\documentclass[doublecol]{epl2} 
\title{Fermi arcs and isotope effect of the
magnetic penetration depth in underdoped cuprates}
\shorttitle{Isotope effect of the
magnetic penetration depth in cuprates} 
\author{Roland Zeyher\inst{1} \and Andr\'{e}s Greco\inst{2}}
\shortauthor{R. Zeyher and A. Greco}

\institute{                    
  \inst{1} Max-Planck-Institut f\"ur Festk\"orperforschung - Heisenbergstrasse 1, D-70569 Stuttgart, Germany\\
  \inst{2} Departamento de F\'{\i}sica, Facultad de Ciencas Exactas e Ingenier\'{\i}a and IFIR(UNR-CONICET) -
Av. Pellegrini 250, 2000 Rosario, Argentina
}

\date{\today}


\abstract{
The isotope coefficient $\beta$ of the magnetic penetration depth 
in the superconducting state is studied at $T=0$
for a $d$-CDW and a nodal metal model. Disregarding 
superconductivity the Fermi surface of the first model possesses arcs  
whereas the second model has no arcs.
We show that a large increase of $\beta$ in the pseudogap region is 
generically incompatible with Fermi arcs in the pseudogap state.  
Thus only the second model shows a large increase of
$\beta$ with decreasing doping. The required electron-phonon coupling is small and 
compatible with first-principles calculations based on the local density 
approximation (LDA). 
}

\begin{document} 
\maketitle

%\section{\label{sec:intro} Introduction}
Underdoped high-$T_c$ cuprates show properties which are not expected to occur 
in Fermi liquids. \cite{Timusk} There is an energy gaplike 
feature (pseudogap) seen already 
well above the transition temperature $T_c$ which increases with decreasing 
doping. At the same time the large Fermi surface of the
normal state at large dopings transforms in the underdoped region into 
arcs around the nodal direction and becomes gapped near the antinodal 
points. \cite{Norman} 
The temperature dependence of the length of these Fermi arcs is presently 
controversely discussed. In particular, it is debated whether for 
$T \rightarrow 0$ and in the absence of  superconductivity their lengths
would approach zero \cite{Kanigel2} or a finite value. \cite{Tallon2} 

Another unexpected feature
of underdoped cuprates is the observed large isotope effect $\beta$ for the 
magnetic penetration depth. $\beta$ is very small in the overdoped region, 
increases strongly with decreasing
doping, and may then reach values of the order of one. \cite{Keller,Tallon} 
Such large values cannot be explained
by Eliashberg theory where $\beta$ is practically zero in agreement with
a recent experiment in the strong-coupling superconductor 
MgB$_2$ \cite{Castro}. 

The aim of this Letter is to show the strong interrelation between 
the pseudogap and Fermi arcs at T=0 and   
$\beta$ in the underdoped regime.  
Anomalously large isotope coefficients due to the presence
of the pseudogap have been suggested previously. \cite{Pringle}
For the coefficient $\alpha$, describing the change in $T_c$ due to isotope 
substitution,
detailed quantitative calculations \cite{Dahm,Zeyher}
yielded large enhancements of $\alpha$ even for a small electron-phonon (EP)
coupling
due to the presence of a pseudogap in rough agreement with the experiments. 
Similar calculations for $\beta$, presented in this Letter, yield a 
more complex picture. An enhancement of $\beta$ does not only require a
pseudogap but it is generally incompatible with Fermi arcs at $T=0$ formed by
infinitely sharp quasiparticles. To show this we first study a  
$d$-CDW model which can be derived from the $t$-$J$ model 
within mean field theory. Such a model is generic for models where the pseudogap 
is associated with long-range order in the particle-hole channel yielding Fermi 
arcs at $T = 0$ due to imperfect nesting.
We show that this model leads to neglegible values for $\beta$
in the underdoped region because some integrals perpendicular to the arcs
diverge if the superconducting gap tends to zero. 
One way to avoid these divergencies is provided by the nodal metal model
which has no arcs. Explicit calculations show in this case indeed 
a large increases of $\beta$ and a ratio of $\beta/\alpha$ of about 2 
in the underdoped region in agreement with experiment.

Our calculation is based on the large $N$ limit of the $t-J$ model 
(N is the number of spin components) \cite{Affleck} which represents a
model with competing $d$-CDW and $d$-wave superconducting order 
parameters. \cite{Cappelluti} 
For calculating $\beta$ we add as in Ref. \cite{Zeyher} a 
phonon-induced interaction $V({\bf k}-{\bf k'})$
between electrons.
In the following only the $d$-wave part of $V$ is required which
is obtained by replacing $V({\bf k}-{\bf k'})$
by $4V \gamma({\bf k})\gamma({\bf k'})$ which defines the $d$-wave 
coupling constant $V$ for a phonon-induced nearest neighbor interaction.
$\gamma({\bf k})$ is equal to $(\cos(k_x)-\cos(k_y))/2$.

The isotope coefficient $\beta$ for the phase 
stiffness $\Lambda$ is defined by
$\beta = \frac{1}{2} \frac{\partial \ln \Lambda}{\partial \ln \omega_D}$.
$\Lambda$ is related to the magnetic penetration depth $\lambda$
by $\Lambda = c^2/(4\pi e^2\lambda^2)$ where $c$ and $e$ are the 
velocity of light and the electronic charge, respectively. 
$\omega_D$ is the Debye frequency which is assumed to be
proportional to $M^{-0.5}$ where $M$ is the ionic mass. At zero
temperature only the diamagnetic term contributes to $\Lambda$ 
in the superconducting state which is
given by 
\begin{equation}
\Lambda = \frac{1}{N_c} \sum_{{\bf k},\alpha}n_\alpha({\bf k})
\frac{\partial^2 \chi_\alpha({\bf k})}{\partial k_x^2}.
\label{LambdaD}
\end{equation}
Here and in the following we use a reduced zone scheme anticipating
that the original high-temperature primitive cell doubles for a $d$-CDW
ground state. Thus the momentum ${\bf k}$
in Eq. (\ref{LambdaD}) runs only over half of the high-temperature 
Brillouin zone whereas the index $\alpha=1,2$ counts the original and
the backfolded electronic branches. According to Peierl's substitution rule
$\chi_\alpha({\bf k})$ are the electronic eigenstates in the normal state
renormalized by many-body interaction within mean-field theory. Thus they may
describe a pseudogap in the particle-hole channel but are unaffected by 
the superconducting gap. In our case they are given by
\begin{equation}
\chi_{1,2}({\bf k}) = \frac{\epsilon_+({\bf k})}{2} \pm \frac{1}{2}
\sqrt{\epsilon^2_-({\bf k}) +4\Phi^2({\bf k})},
\label{eigen}
\end{equation}
with $\epsilon_{\pm}({\bf k}) = \epsilon({\bf k})
\pm \epsilon({\bf k}-{\bf Q})$, where ${\bf Q} = (\pi,\pi)$ is the wave vector of
the $d$-CDW. $\Phi({\bf k})$ is the amplitude of the $d$-CDW 
and equal to $\Phi_0 \gamma({\bf k})$ with $\gamma({\bf k})=(\cos(k_x)-\cos(k_y))/2$. The bare
electronic energies $\epsilon({\bf k})$ are those of the
$t$-$J$ model in the large $N$ limit counted from the chemical potential $\mu$,
i.e.,
$\epsilon({\bf k}) = -2(\delta t + r J)(\cos(k_x)+\cos(k_y))
-4t'\delta \cos(k_x)\cos(k_y)  -\mu$,
with $r = 1/N_c \sum_{\bf q} \cos({q_x})f(\epsilon({\bf q}))$.
$f$ is the Fermi function, $\delta$ the doping away from half-filling, $J$
the Heisenberg coupling constant and $t$ and $t'$ are hopping amplitudes
between nearest and next nearest neighbors on a square lattice, respectively.
$N_c$ is the original number of 
primitive cells. $n_\alpha({\bf k})$ denotes the electron density with momentum
${\bf k}$ and is given by
\begin{equation}
n_\alpha({\bf k}) = 2T \sum_n G_{11}^{(\alpha)}({\bf k},i\omega_n)
e^{i\omega_n\eta},
\label{n}
\end{equation}
where the prefactor 2 accounts for the spin degeneracy,
$\eta$ is an infinitesimally small positive quantity and $\omega_n$
the Matusbara frequency $(2n+1)\pi T$. 
$G_{11}^{(\alpha)}$ is the element (1,1) of the Green's function 
matrix $G^{(\alpha)}({\bf k},i\omega_n)$ defined by its inverse as, 
\begin{equation}
 {G^{-1}}^{(\alpha)}({\bf k},i\omega_n)=
\left( 
\begin{array}{c c}  
i\omega_n -\chi_\alpha({\bf k}) & -\Delta({\bf k},n)  
                \\
-\Delta({\bf k},n) & i\omega_n +\chi_\alpha({\bf k})  

\end{array} \right).
\label{matrix}
\end{equation}
The superconducting gap function $\Delta({\bf k},n)$ consists of
a contribution due to the $d$-wave part of the Heisenberg and one due to
the $d$-wave part of the EP interaction,
\begin{equation}
\Delta({\bf k},n) = -\Bigl(\sqrt{J} \Delta_1 + \sqrt{V}
\Theta(\omega_D-|\omega_n|)\Delta_3\Bigr)
\sqrt{2}\gamma({\bf k}),
\label{Delta}
\end{equation}
where  $\Theta$ is the step function. 
In Eqs. (\ref{matrix})-(\ref{Delta})
we used the fact that the phonon-induced reduction of the quasi-particle
weight is independent of the ionic mass \cite{Zeyher} and thus may be
neglected in the following.
  
The self-consistency equation for $\Delta$
splits up into two equations for $\Delta_1$ and $\Delta_3$, reading
\begin{equation}
(1+F_{11}) \Delta_1 + F_{12} \Delta_3
=0,
\label{F1}
\end{equation}
\begin{equation}
F_{12} \Delta_1 +(1+F_{22}) \Delta_3 
=0,
\label{F2}
\end{equation}
with
\begin{equation}
F_{11}(\Delta_1,\Delta_3,\omega_D) 
= \frac{2JT}{N_c} \sum_{{\bf k},n,\alpha} \frac{\gamma^2({\bf k})}
{(i \omega_n)^2-\chi^2_\alpha({\bf k})-\Delta^2({\bf k},n)},
\label{F11}
\end{equation}
\begin{eqnarray}
F_{12}(\Delta_1,\Delta_3,\omega_D) =  
\frac{2\sqrt{JV}T}{N_c} \cdot \hspace*{3.7cm} \nonumber \\  
\sum_{{\bf k},n,\alpha}
\frac{\gamma^2({\bf k}) \Theta(\omega_D-|\omega_n|)}
{(i \omega_n)^2-\chi^2_\alpha({\bf k})-\Delta^2({\bf k},n)}.
\label{F12}
\end{eqnarray}
$F_{22}$ is given by the expression for $F_{12}$ if one replaces there
$\sqrt{JV}$ by $V$.

%%%%%%%%%%%%%%%%%%%%%%%%%%%%%%%% FIG. 1 %%%%%%%%%%%%%%%%%%%%%%%%%%%%%%%%%%%%
\begin{figure}[t] 
\vspace*{-1.4ex}
\includegraphics[angle=270,width=8.0cm]{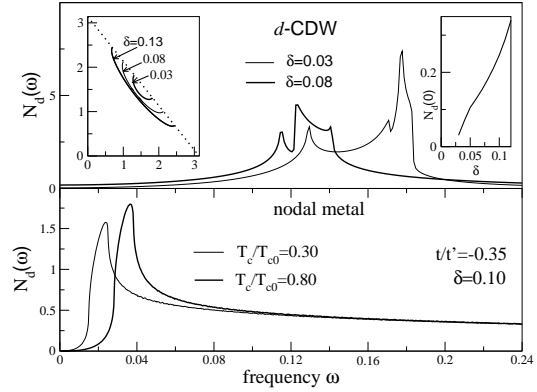}
\caption{\label{fig:1}
$D$-wave projected density of states $N_d(\omega)$ in the pseudogap state 
without superconductivity for the the $d$-CDW (upper diagram)
and the nodal metal model (lower diagram) at $T=0$.
Left and right insets show Fermi lines and $N_d(0)$, respectively, 
at different dopings.
}
\end{figure}
%%%%%%%%%%%%%%%%%%%%%%%%%%%%%%%%%%%%%%%%%%%%%%%%%%%%%%%%%%%%%%%%%%%%%%%%%%%%

One important ingredient of the theory is the $d$-wave projected density
$N_d(\omega)$ of the $d$-CDW model. It is shown in the upper diagram of Fig. \ref{fig:1} 
together with Fermi lines in the left inset. 
For the calculation we took the values $J=0.3$ and $t'=-0.35$
measuring all energies in units of $t$.
The right inset in Fig. \ref{fig:1} shows
that $N_d(0)$ is nonzero at finite $\delta$ due to finite 
arcs and approaches zero only in the limit $\delta \rightarrow 0$.

Recent LDA calculations \cite{Heid,Giustino} showed that the 
EP interation and in particular its $d$-wave part 
\cite{Heid1} is very small in cuprates. This allows to simplify Eqs.(\ref{F11})
and (\ref{F12}) by keeping only the leading terms in $V$. In a first step
one considers $V=0$ so that $\Delta_3 =0$ 
and determines in the $d$-CDW case self-consistently 
$\Delta_1 = \bar{\Delta}_1$ and $\Phi({\bf k})$
similar as in Ref. \cite{Zeyher}. Writing then
$\Delta_1 = \bar{\Delta}_1 + \Delta_2$    
one may neglect $F_{22}$ in
Eq.(\ref{F2}) and obtains from this equation
\begin{equation}
\Delta_3 = -F_{12}(\bar{\Delta}_1,0,\omega_D)\bar{\Delta}_1.
\label{beta}
\end{equation}
Eq.(\ref{F11}) yields, again in leading order in $V$,
\begin{eqnarray}
\Delta_2 = \Bigl(F^2_{12}(\bar{\Delta}_1,0,\omega_D)-\Delta_3\;
\frac{\partial F_{11}(\bar{\Delta}_1,\Delta_3,\omega_D)}{\partial \Delta_3}\Bigr)
/  \nonumber \\
\frac{\partial F_{11}(\Delta_1,0,\omega_D)}{\partial \Delta_1}.
\label{alpha}
\end{eqnarray}
After taking the derivatives in Eq. (\ref{alpha}) 
$\Delta_1$ and $\Delta_3$ should be put to $\bar{\Delta}_1$ and 0, respectively.
Assuming that $\Lambda$ depends on $\omega_D$ only via
$\Delta_1$ and $\Delta_3$ and noting that Eqs. (\ref{beta}) and (\ref{alpha})
hold for a general $\omega_D$ one may easily form the derivatives of
$\Delta_2$ and $\Delta_3$ with respect to $\omega_D$ and thus 
obtain $\beta$.
%%%%%%%%%%%%%%%%%%%%%%%%%%%%%%%% FIG. 2 %%%%%%%%%%%%%%%%%%%%%%%%%%%%%%%%%%%%
\begin{figure}[t] 
\vspace*{-0.5cm}
\includegraphics[angle=270,width=8.cm]{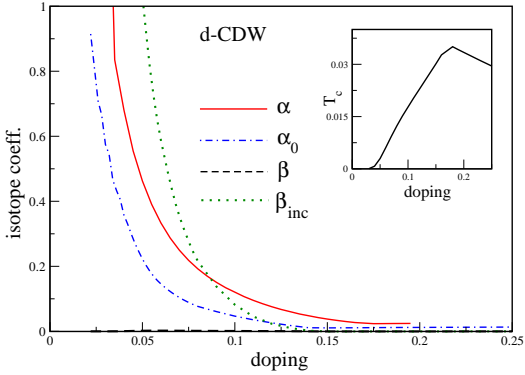}
\caption{\label{fig:2}
(Color online)
Isotope coefficients $\alpha$, $\alpha_0$, $\beta$, and $\beta_{inc}$
for the $d$-CDW model as a function of doping using $V=0.04$
and $\omega_D = 0.16$. Inset: Suppression of $T_c$ as a function of doping for the
$d$-CDW model due to the evolving pseudogap. \cite{Zeyher}  
}
\end{figure}
%%%%%%%%%%%%%%%%%%%%%%%%%%%%%%%%%%%%%%%%%%%%%%%%%%%%%%%%%%%%%%%%%%%%%%%%%%%%

In principle there are two more sources for a 
dependence of $\Lambda$
on $M$: The bare one-particle energies $\epsilon({\bf k})$ may
depend on $M$. Such a case may occur if the EP interaction can no longer
be described by Eliashberg theory because strong polaronic or 
non-adiabatic effects play an important role. However, there is overwhelming
evidence, both from experiment \cite{Dessau} and 
theory \cite{Heid,Giustino}, that this is not the case. For instance,
recent angle resolved photoemission (ARPES) experiments on nearly optimally 
doped Bi2212 \cite{Dessau} show that the band width
does not change within the experimental error when substituting 
O$^{16}$ by O$^{18}$. Using  the polaronic model \cite{Alexandrov} an observed value 
of $\beta \sim 1$ would require a 
12 per cent change of the electronic band width which is not observed.  
Another possibility for a $M$ dependence of $\Lambda$ could be due to
the $d$-CDW gap $\Phi({\bf k})$. Recently we have shown \cite{Zeyher} 
that the onset 
temperature $T^\ast$ for the $d$-CDW gap shows practically no isotope effect
which should also hold for the $T=0$ $d$-CDW gap $\Phi({\bf k})$.   

Numerical results for the isotope coefficients $\alpha$, $\alpha_0$, and $\beta$, 
which are related to $T_c$, the $T=0$ superconducting gap and the phase stiffness, 
respectively,
are shown in Fig. \ref{fig:2} for the $d$-CDW model. The calculation of $\alpha$ has been
described previously \cite{Zeyher}, $\alpha_0$ is defined by 
\begin{equation}
\alpha_0 = \frac{\omega_D}{2\Delta_1}\frac{\partial \Delta_1}{\partial \omega_D}.
\label{alphaa0}
\end{equation}
Outside of the pseudogap state, i.e., for $\delta \geq \delta_c \sim 0.145$,
all three coefficients are very small. Here $\alpha$ and $\alpha_0$
are much smaller than the BCS value of 1/2 because superconductivity is mainly 
determined by $J$ and not by $V$ because of the small employed value for 
$V$. Entering the pseudogap state
$\alpha$ and $\alpha_0$ nevertheless increase strongly with decreasing doping
due to the evolving pseudogap as explained in detail in Ref. \cite{Zeyher}.
In contrast to that $\beta$ does not change much below $\delta_c$ and remains
practically zero throughout the underdoped region. Considering only the underdoped
region, i.e., $\delta \leq \delta_c$, one could also use the ratio $T_c/T_{c0}$
as an independent variable instead of $\delta$. The curves in Fig. \ref{fig:2} would not
look much different in such a plot because $T_c$ depends rather linearly
on $\delta$ in that region, as shown in the inset of Fig. \ref{fig:2}.

To understand the surprising behavior of $\beta$ we present now approximate 
analytic expressions for $\beta$ and various other quantities which become
asymptotically exact at small values of the superconducting gap but are also 
rather accurate over the whole doping region.
Evaluating Eq.(\ref{LambdaD}) for an infinite cutoff and neglecting momentum
derivatives of the order parameter we find
\begin{equation}
\Lambda = \frac{1}{2N_c}\sum_{{\bf k},\alpha}\frac{\Delta^2({\bf k})
(\nabla \chi_\alpha({\bf k}))^2}{E_\alpha^3({\bf k})},
\label{Lam}
\end{equation}
and 
\begin{equation}
X=(\frac{\beta}{\alpha_0}+1)\frac{\Lambda}{3} =
\frac{1}{2N_c}\sum_{{\bf k},\alpha}\frac{\Delta^2({\bf k})\chi^2_\alpha({\bf k})
(\nabla \chi_\alpha({\bf k}))^2}{E_\alpha^5({\bf k})}.
\label{X}
\end{equation}
The main contribution in the sums over momentum in Eqs.(\ref{Lam}) and (\ref{X})
comes from the region near the arcs which consists of all points $\{{\bf k}_{F\alpha}\}$
satisfying $\chi({\bf k}_{F\alpha}) = 0$. Correspondingly, we split these sums into a
part parallel and a part perpendicular to the arcs. Assuming that the electron dispersion
perpendicular to the arcs can be approximated linearly we find that the 
integrations perpendicular to the arcs diverge for small $\Delta({\bf k})$
cancelling, for instance, the prefactor $\Delta^2({\bf k})$ in Eq.(\ref{Lam}).
The remaining integration parallel to the arcs in $\Lambda$ and $X$ can be written
as surface integrals along the arcs yielding
\begin{equation}
\Lambda = \sum_\alpha \oint \frac{dS({\bf k}_{F\alpha})}
{4\pi^2v({\bf k}_{F\alpha})} (\nabla \chi_\alpha({\bf k}_{F\alpha}))^2,
\label{int}
\end{equation} 
and $X = \Lambda/3$, or equivalently, 
\begin{equation}
\beta/\alpha_0 = 0.
\label{main} 
\end{equation}
Eq.(\ref{main}) explains the tiny values for $\beta$ in Fig. \ref{fig:2} obtained
by numerical evaluation of the exact expressions for $\beta$. The small values
for $\beta$ throughout the underdoped region reflect directly the divergencies
encountered in the momentum integrations perpendicular to the arcs. If there would
be no divergencies the right-hand side of Eq.(\ref{X}) would smoothly approach
the right-hand side of Eq.(\ref{Lam}) for $\Delta({\bf k}) \rightarrow 0$. Thus
we would obtain $\beta/\alpha_0 = 2$ and an increasing $\beta$ towards
small dopings both of which would be in excellent agreement with experiment.
This suggests that the above mentionned divergencies actually do not occur
in the high-$T_c$ superconductors so far studied with respect to $\beta$.

One model which has a pseudogap but no arcs and thus no divergencies in
momentum sums is the nodal metal (NM) model \cite{Kanigel2,Dahm}.
Its dispersion is given by
\begin{equation}
\chi_{1,2}({\bf k}) = \pm 
\sqrt{\epsilon^2({\bf k}) +\Phi^2({\bf k})},
\label{eigen1}
\end{equation}
which formally can be obtained from Eq. (\ref{eigen}) by putting
$\epsilon_+({\bf k})=0$. The NM model does not allow to determine
$\Phi_0$ but considers it as a parameter. Following Ref. \cite{Dahm}
it is convenient not to use $\Phi_0$ as an independent variable but the 
ratio $T_c/T_{c0}$ which describes the reduction of $T_c$ due to $\Phi({\bf k})$
relative to the maximum transition temperature $T_{c0}$ at $\delta = \delta_c$.
Dependencies on  $T_c/T_{c0}$ may be interpreted
as doping dependencies due to the monotonic relation between $T_c$ and $\delta$
in the underdoped region, see the inset of Fig. \ref{fig:2} in case of the $d$-CDW system. 
The lower diagram in Fig. \ref{fig:1} shows $N_d(\omega)$
for the NM model for $\delta = 0.10$ and two values for $T_c/T_{c0}$.
$N_d(0)$ is zero in all cases because of the absence of arcs.

\begin{figure}[t] 
\vspace*{-0.5cm}
\includegraphics[angle=270,width=8.0cm]{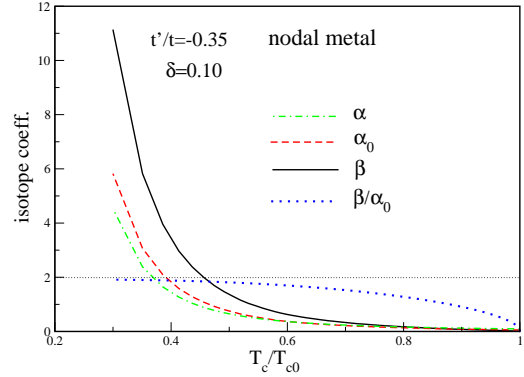}
\caption{\label{fig:3}
(Color online)
Isotope coefficients $\alpha$, $\alpha_0$, $\beta$ and the ratio $\beta/\alpha_0$
for the nodal metal model as a function of $T_c/T_{c0}$ for a fixed doping $\delta = 0.10$
using $V=0.04$ and $\omega_D=0.16$.
}
\end{figure}

The calculation of $\beta$ proceeds in the same way as for the $d$-CDW model.
Fig. \ref{fig:3} shows numerical results for $\alpha, \alpha_0, \beta,$ and the 
ratio $\beta/\alpha_0$ using $V=0.04$ and $\omega_D=0.16$. Unlike in Fig. \ref{fig:2}
$\beta$ increases strongly with decreasing $T_c/T_{c0}$ and the ratio $\beta/\alpha_0$
is near two over a large part of the underdoped region, approaching exactly 2
for  $T_c/T_{c0} \rightarrow 0$. Since $\alpha$ and $\alpha_0$ are of similar
magnitude this implies also $\beta/\alpha \sim 2$ which is exactly the value found in
experiment. The diagram also shows that $\beta$ assumes values of 1 and larger already 
for moderate reductions in $T_c$. This is rather astonishing because it means that
the small EP coupling constant found in LDA calculations \cite{Heid,Giustino,Heid1}
is able to produce the large value for $\beta$ of Fig. \ref{fig:3}.
Varying the Debye frequency over a large frequency region 
does not change much the curves in Fig. \ref{fig:2} and Fig. \ref{fig:3}.
Similar as in the case of $\alpha$ discussed in Ref. \cite{Zeyher}
$\alpha_0$ and $\beta$ do not show anomalies if the phonon frequency 
and the pseudogap are close to each other. 
Actually, for all phonon frequencies smaller than about 0.16 the
pseudogap crosses the phonon at some doping without any effect on 
the curves showing that the large increases in the underdoped
region are not due to resonance effects between the phonon and the
pseudogap. 

Important details of Fig. \ref{fig:3} such as the behavior of the ratio $\beta/\alpha_0$
can be understood again by an approximate analytic evaluation of the sums over
momenta. Since the NM model has no arcs one may expand $\epsilon({\bf k})$
around the high-temperature Fermi line given by $\epsilon({\bf k}) = 0$. 
Splitting again the momenta sums into perpendicular and parallel parts,
taking $\Phi({\bf k})$ right on the Fermi line and linearising $\epsilon({\bf k})$
the integration perpendicular to the Fermi line can be carried out and is
always finite. As a result $\beta$ and $X$ can again be written as surface
integrals. Due to the absence of divergencies $X \rightarrow \Lambda$
for $\Delta({\bf k}) \rightarrow 0$ yielding $\beta/\alpha_0 \rightarrow 2$.
This result is rather general and just expresses the absence of divergencies. 

%%%%%%%%%%%%%%%%%%%%%%%%%%%%%%%% FIG. 3 %%%%%%%%%%%%%%%%%%%%%%%%%%%%%%%%%%%%
%%%%%%%%%%%%%%%%%%%%%%%%%%%%%%%%%%%%%%%%%%%%%%%%%%%%%%%%%%%%%%%%%%%%%%%%%%%%

Many features of the NM model agree with the experiment, for instance,
the observed ratio
$\beta/\alpha \sim 2$ over most of the underdoped region and 
the large increase of $\alpha$ and $\beta$ with decreasing $T_c$. 
Obviously it has also short-comings. For instance, the lower
diagram of Fig. 1 implies a pseudogap of only about 0.02 or 10 meV for
a decrease $T_c/T_{co}$ of 0.3 which is unrealistically small.
The more fundamental question, however, is whether this model 
with perfect nesting 
in the particle-hole channel is applicable to the cuprates at all.
ARPES experiments in extremly underdoped Bi2212 \cite{Chatterjee} 
found an excitation spectrum 
of the pseudogap phase without superconductivity which is very
similar to the NM model. These data, however, have been interpreted
in terms of a superconducting order parameter with finite
phase correlation length. Since superconductivity does not renormalize
the one-electron spectrum entering the phase stiffness the resulting
$\beta$ would be equally small as in a superconductor without pseudogap
and thus disagree with experiment. Nodal metal behavior can, however,
be realized to some extent for an incommensurate $d$-CDW state.
The nesting condition $\epsilon_+({\bf k}) = 0$ of the NM model
can in this case fulfilled in limited regions around ${\bf k}$ points
on Fermi lines which are connected by nesting vectors and exhibit parallel 
tangents to the Fermi line. Assuming four nesting vectors of the form
${\bf Q} = (\pi, \pi \pm \delta_{inc})$ and $(\pi \pm \delta_{inc}, \pi)$, 
determining $\delta_{inc}$
from the nesting condition and correlating only pairs of states with
maximal nesting we obtained the dotted line $\beta_{inc}$ in Fig. 2 for $\beta$.
It shows the desired increase towards low doping but the calculated
ratio $\beta_{inc}/\alpha_{0,inc}$ (not shown in Fig. 2) deviates
substantially from 2 due to divergencies at Fermi lines near the 
antinodal point. It also has been shown \cite{Zeyher1} that superconductivity
(not taken into account above in determining $\delta_{inc}$)
suppresses $\delta_{inc}$ strongly, suggesting that incommensurability
is not causal for the large observed values for $\beta$.  

The above results show that the presence of arcs (or, more general,
a finite length of the Fermi line) at $T=0$ in the pseudogap
state (i.e., without superconducitvity) are generically   
incompatible with large values of $\beta$ in the underdoped region.   
This holds for all models where the pseudogap phase has long-range
order and arises in the particle-hole channel which necessarily leads to
arcs in two-dimensional models due to imperfect nesting. The $d$-CDW
model is one example. It also holds for short-range models \cite{Kampf}
for the pseudogap which are Fermi liquids, i.e., where the self-energy 
at $T=0$ can expanded in powers of the momentum and frequency. In this case the arcs
are nothing else than the Fermi lines in the normal state without pseudogap 
and the arguments used above for Fermi arcs apply. 
Large values for $\beta$ can therefore only expected if no arcs
exist at all as in the NM model or that the divergencies in the
integration perpendicular to the arcs in calculatling $\beta$ are not present for some reason.
One possibility could be a non-Fermi liquid ground state without
infinitely sharp quasiparticles and a broad spectral function.  

In conclusion we have investigated the
isotope coefficient $\beta$ of the magnetic penetration depth   
at $T=0$ using two models for the pseudogap. The calculation shows
that $\beta$ depends severely on the presence or absence of Fermi arcs
at $T=0$ in the absence of superconductivity. The experimentally
observed large increase of $\beta$ in the underdoped region is
reproduced in the case of a nodal metal using a small EP
coupling consistent with first-principles calculations based on the
local density approximation. The $d$-CDW model is generic for models
with two competing order parameters often used to interpret experimental
results. \cite{Ma,Kondo} The resulting $\beta$, however, is small throughout
the underdoped region and disagrees with experiment even if the EP
coupling constant is increased by orders of magnitude.
Responsible for the absence of an enhancement of $\beta$ is the
presence of Fermi arcs which lead to divergent ${\bf k}$ integrals 
perpendicular to the arcs in calculating $\beta$. 
We hope that the discovered link between
arcs and the isotope coefficient $\beta$ motivates
more experimental work on the doping dependence of $\beta$.

The authors thank D. Manske for a critical reading of the manuscript.
R.Z. and A.G. are grateful to the Dep. de F\'{\i}sica (Rosario) and
the MPI-FKF (Stuttgart), respectively, for hospitality and financial support.
%%%%%%%%%%%%%%%%%%%%%%%%%%%%%%%%%%%%%%%%%%%%%%%%%%%%%%%%%%%%%%%%%%%%%%%%%%%%
%%
%%                           REFERENCES
%%
%%%%%%%%%%%%%%%%%%%%%%%%%%%%%%%%%%%%%%%%%%%%%%%%%%%%%%%%%%%%%%%%%%%%%%%%%%%%

\end{document}